\documentclass[aps,prl,twocolumn,superscriptaddress,showpacs]{revtex4-1}

\usepackage{graphicx}
\usepackage{amsmath}
\usepackage{color}
\begin{document}


\title{First Principles explanation of the positive Seebeck coefficient of lithium}


\author{Bin Xu}
\email[Author to whom correspondence should be addressed. Email address: ]{bxu@ulg.ac.be}
\author{Matthieu J. Verstraete}
\affiliation{D\'{e}partment de Physique, Universit\'{e} de Li\`{e}ge, All\'{e}e du 6 Ao\^ut 17, B-4000 Sart Tilman, Belgium}
\affiliation{European Theoretical Spectroscopy Facility (http://www.etsf.eu)}


\date{\today}

\begin{abstract}
  Lithium is one of the simplest metals, with negative charge carriers and a close reproduction of free electron dispersion. Experimentally, however, Li is one of a handful of elemental solids (along with Cu, Ag, Au etc.) where the sign of the Seebeck coefficient ($S$) is opposite to that of the carrier. This counterintuitive behavior still lacks a satisfactory interpretation. We calculate $S$ fully from first-principles, within the framework of P.B. Allen's formulation of Boltzmann transport theory. Here it is crucial to avoid the constant relaxation time approximation, which gives a sign for $S$ which is necessarily that of the carriers. Our calculated $S$ are in excellent agreement with experimental data, up to the melting point. In comparison with another alkali metal Na, we demonstrate that within the simplest non-trivial model for the energy dependency of the electron lifetimes, the rapidly increasing density of states (DOS) across the Fermi energy is related to the sign of $S$ in Li. The exceptional energy dependence of the DOS is beyond the free-electron model, as the dispersion is distorted by the Brillouin Zone edge, a stronger effect in Li than other Aliki metals. The electron lifetime dependency on energy is central, but the details of the electron-phonon interaction are found to be less important, contrary to what has been believed for several decades. Band engineering combined with the mechanism exposed here may open the door to new ``ambipolar'' thermoelectric materials, with a tunable sign for the thermopower even if either n- or p-type doping is impossible.
\end{abstract}



\maketitle



%

Thermoelectricity (TE) has drawn much attention over the past century\cite{zebarjadi, snyder_2008_thermoelectrics} as an effective way of producing electricity from heat energy, or vice versa. In addition to applications in waste heat recovery, the reversible functionality of TE materials also enables heating and refrigeration within the same unit. Spot cooling\cite{wang:034503} of computer processors can be achieved with TE devices of small size and without moving parts. The efficiency of current thermoelectric devices is relatively low compared to, e.g., thermal engines, which limits their applications.\cite{zebarjadi} In the search for a good thermoelectric material, a large Seebeck coefficient ($S$) is one of the central components in the figure of merit, where it appears squared. 
Most advances in TE have however targeted the simpler tasks of lowering the thermal conductivity,\cite{biswas_2012_all_scale_nanostructuring} or optimizing the electron density of states\cite{pei_2011_band_effects_thermoelectrics}.
The magnitude of $S$ is also important in other applications, e.g. for thermal sensors.\cite{van1986} 
Though $S$ can be measured straightforwardly in experiment and calculated theoretically within certain approximations, a complete microscopic understanding, and paths for systematic improvement of $S$ are still lacking. The most common approach is to consider a constant averaged relaxation time for the electrons ($\tau$). The relaxation time approximation (RTA) works in a surprisingly large number of cases, but has little formal justification, and we expose some more of its limitations below.

Materials with n-type carriers should yield negative $S$, as electrons diffuse from the high temperature side to the low temperature side. For the monovalent metal Li at ambient pressure this is not the case. Lithium exhibits positive $S$ from low to high temperatures, through a martensitic transformation at 77 K\cite{vaks1989} and melting at 454 K\cite{bidwell,kendall,surla,LBdatabase}. This is in contrast to most simple metals, in particular other Alkali metals. An explanation was proposed by Robinson nearly half a century ago based on a nearly-free-electron model, where the positive Seebeck of Li was attributed to the energy variance of the mean free path around the Fermi energy\cite{robinson1967}, due to an unusual energy dependence of the electron-phonon interaction\cite{robinson1968}. These calculations adopted model interactions for scattering of electrons by lattice vibrations, which relied on empirical parameters. On the other hand, a controversial argument was proposed by Jones in the 1950s, that the positive Seebeck in monovalent metals is due to the significant deviation from the free-electron density of states (DOS) as part of the Fermi surface lies close to the Brillouin zone boundary.\cite{jones1955} No actual calculation was carried out to verify this hypothesis. 
In this Letter, we revisit the anomalous sign of $S$ in Li through fully \textit{ab initio} calculations within the framework of Boltzmann's transport theory. To understand the positive sign of $S$ in Li, we also explore the $S$ of sodium metal, for a comparative analysis. Contrary to Robinson's hypothesis, while consistently with Jones', we indeed find significant deviation from the free-electron model in Li, and that the electron-phonon interaction is not primarily responsible for the positive $S$. Furthermore we show that, within a simple model for the energy dependent lifetime $\tau(\epsilon)$, the energy projected conductivity (a purely electronic quantity) determines the sign of $S$.


To calculate the Seebeck coefficient, we adopt the lowest-order variational approximation (LOVA) to the Boltzmann transport equation\cite{allen1978}. For $S$ it is crucial to include explicitly inelastic contributions and Fermi smearing effects. These are beyond the commonly used elastic version of the LOVA\cite{allen1978, savrasov1996}, which leads to $S = 0$. In the LOVA:
\begin{equation} \label{eq:s}
  S_{\alpha \beta}=\frac{\pi k_{\text{B}}}{\sqrt{3}e} \sum_\gamma (Q_{01})_{\alpha \gamma} (Q_{11}^{-1})_{\gamma \beta}
\end{equation} 
where $(Q_{nn'})_{\alpha \beta}$ is the scattering operator for cartesian directions ($\alpha, \beta = x, y, z$), expressed in Allen's basis set (indices $n$, $n'$) and $e$ is the absolute value of the electron charge. The basis set is separated into $\textbf{k}$ dependent ``Fermi-surface harmonics'' (FSH) and energy dependent polynomials (see below). In the LOVA one uses only the lowest non-zero FSH, which is simply a normalized Fermi velocity
viz. $F_\alpha(\textbf{k}) = v_{\alpha}(\textbf{k})/v_{\alpha}(\epsilon_{\text{F}})$.
The normalization is given by $v^2_{\alpha}(\epsilon)=  \left[ \sum_{\textbf{k}} v_{\alpha}^2(\textbf{k}) \delta(\epsilon_{\textbf{k}}-\epsilon)\right] / N(\epsilon)$ where $N(\epsilon)$ is the density of electronic states. 

The scattering operators are calculated as:
\begin{eqnarray} \label{eq:q}
  && (Q_{nn'})_{\alpha \beta}=\frac{2\pi V_{\text{cell}}N(\epsilon_{\text{F}})}{\hbar k_{\text{B}}T} \int d\epsilon  d\epsilon'  d\omega \sum_{s,s'=\pm1} f(\epsilon)\left[ 1-f(\epsilon')\right] \nonumber \\
  && \times \;\; \{ \left[ N(\omega)+1\right] \delta(\epsilon-\epsilon'-\hbar \omega)+N(\omega)\delta(\epsilon-\epsilon'+\hbar \omega) \} \nonumber \\ 
   &&  \times \;\; \alpha_{\text{tr}}^2F(s,s',\alpha,\beta,\epsilon,\epsilon',\omega) J(s,s',n,n',\epsilon,\epsilon')  
\end{eqnarray} 
where $\epsilon, \epsilon'$ are electron energies relative to the Fermi level $\epsilon_{\text{F}}$, $\omega$ is a phonon frequency, $V_{\text{cell}}$ is the unit cell volume, $f$ and $N$ are the Fermi-Dirac and Bose-Einstein distributions at temperature $T$, respectively.
The transport spectral function $\alpha_{\text{tr}}^2F$ is analogous to the Eliashberg spectral function for superconductivity, but weighted by contributions from electron velocities. Among all the mechanisms that affect the electronic transport, here we only consider the electron-phonon coupling (EPC), which is dominant for most materials except at very low temperatures. See the Supplementary Information for definitions and an overview of Allen's formalism.
For the sign of $S$ a crucial quantity is the joint function $J(s,s',n,n',\epsilon,\epsilon')$ in Eq. (\ref{eq:q}):
\begin{equation} \label{eq:j}
\frac{1}{4} \left[ \frac{\zeta_n(\epsilon)}{N(\epsilon)v(\epsilon)} + s \frac{\zeta_n(\epsilon')}{N(\epsilon')v(\epsilon')} \right] 
                \left[ \frac{\zeta_{n'}(\epsilon)}{N(\epsilon)v(\epsilon)} + s' \frac{\zeta_{n'} (\epsilon')}{N(\epsilon')v(\epsilon')} \right]
                \nonumber
\end{equation}
composed of energy polynomials $\zeta_n(\epsilon)$, with $\zeta_0=1$ and $\zeta_1=\sqrt{3}\epsilon/\pi k_{\text{B}}T$.

The EPC matrix elements, phonons, and electronic velocities are calculated within density functional perturbation theory (DFPT),\cite{gonze1997first,gonze1997} carried out using the ABINIT package.\cite{gonze2005} 
The exchange and correlation functional is treated with the local density approximation (LDA). For bcc Li, an unshifted 36$\times$36$\times$36 $\textbf{k}$-point grid and 12$\times$12$\times$12 $\textbf{q}$-point grid are employed, ensuring good convergence for transport properties. For bcc Na, an unshifted $\textbf{k}$-grid of 24$\times$24$\times$24 is found to be sufficient for our comparisons.  For the ground state and DFPT calculations a ``cold smearing'' function \cite{marzari1999} of width 0.04 Ha is used to improve k-grid convergence. The plane-wave basis functions with kinetic energies up to 20 Hartree are used in both systems.

For comparison, the Seebeck coefficient is also calculated within the constant relaxation time approximation, using the BoltzTraP code\cite{madsen}. This approach has often been adopted in theoretical studies of thermoelectric properties. In the RTA

\begin{equation} \label{eq:sbk}
   S_{\alpha \beta}=-\frac{1}{eT} \frac{\int \sigma_{\alpha \beta}(\epsilon) (\epsilon-\epsilon_{\text{F}}) \left(-\frac{\partial f}{\partial \epsilon}\right) d\epsilon}{\int \sigma_{\alpha \beta}(\epsilon) \left(-\frac{\partial f}{\partial \epsilon}\right) d\epsilon}
\end{equation} 
where $\sigma_{\alpha \beta}(\epsilon)$ is $e^2 \tau \sum_{\textbf{k}} v_{\alpha}(\textbf{k})v_{\beta}(\textbf{k}) \delta(\epsilon-\epsilon_{\textbf{k}})$, with $\tau$ the constant relaxation time, which cancels out in Eq. (\ref{eq:sbk}).

\begin{figure}
\includegraphics[width=1.0\columnwidth]{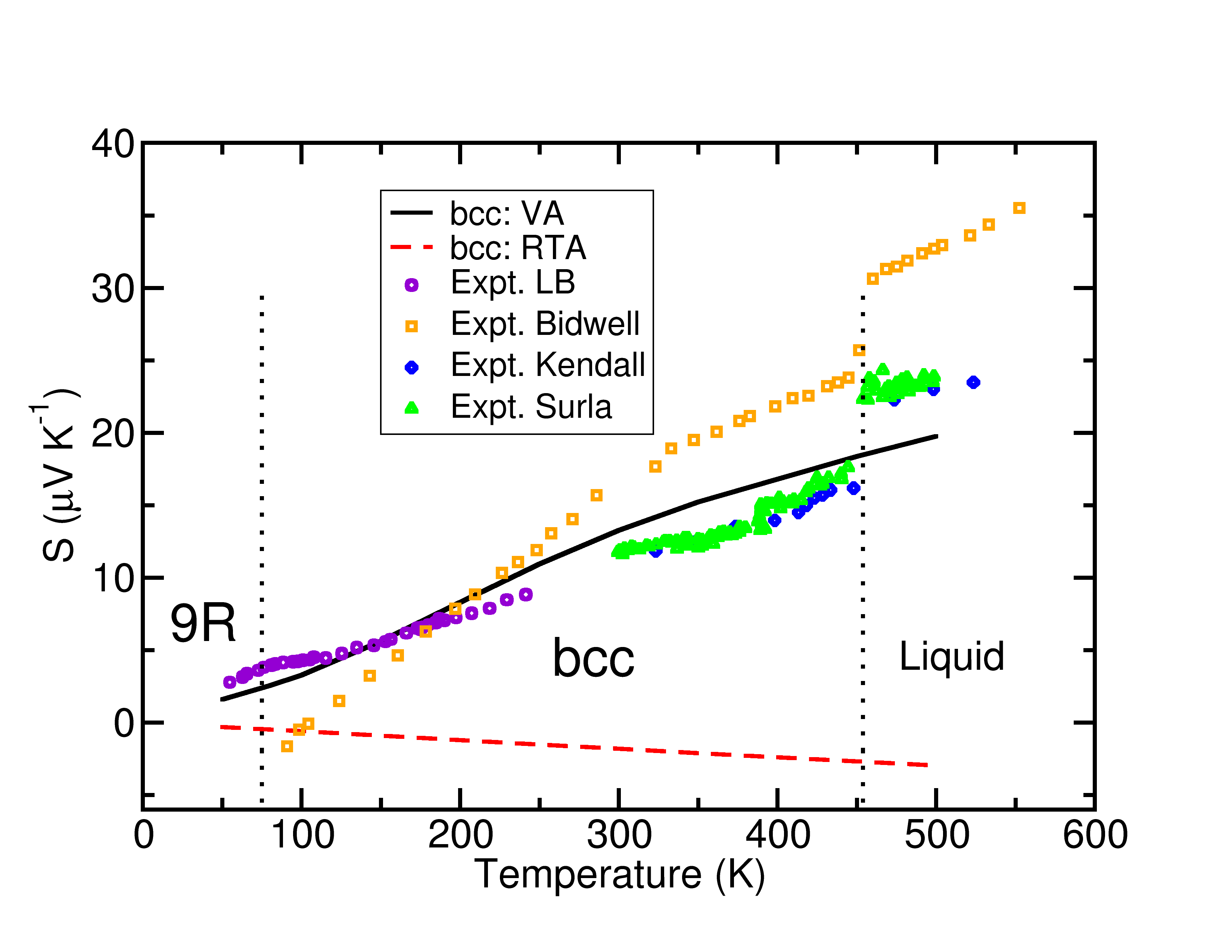}
\caption{(Color online) Seebeck coefficient of Li, as a function of temperature. Black solid line is the calculated $S$ of bcc Li using the variational approach, and the red dashed line denotes calculated result of bcc Li with constant $\tau$ from the BoltrTraP code. Discrete points are experimental data from LB(Landolt-B\"{o}rnstein Database)\cite{LBdatabase}, Bidwell\cite{bidwell}, Kendall\cite{kendall} and Surla \textit{et al.}\cite{surla}. The vertical dotted lines denote the temperatures of 9R-to-bcc phase transition and melting point.} \label{fig:sbk_li}
\end{figure}

\begin{figure}
\includegraphics[width=1.0\columnwidth]{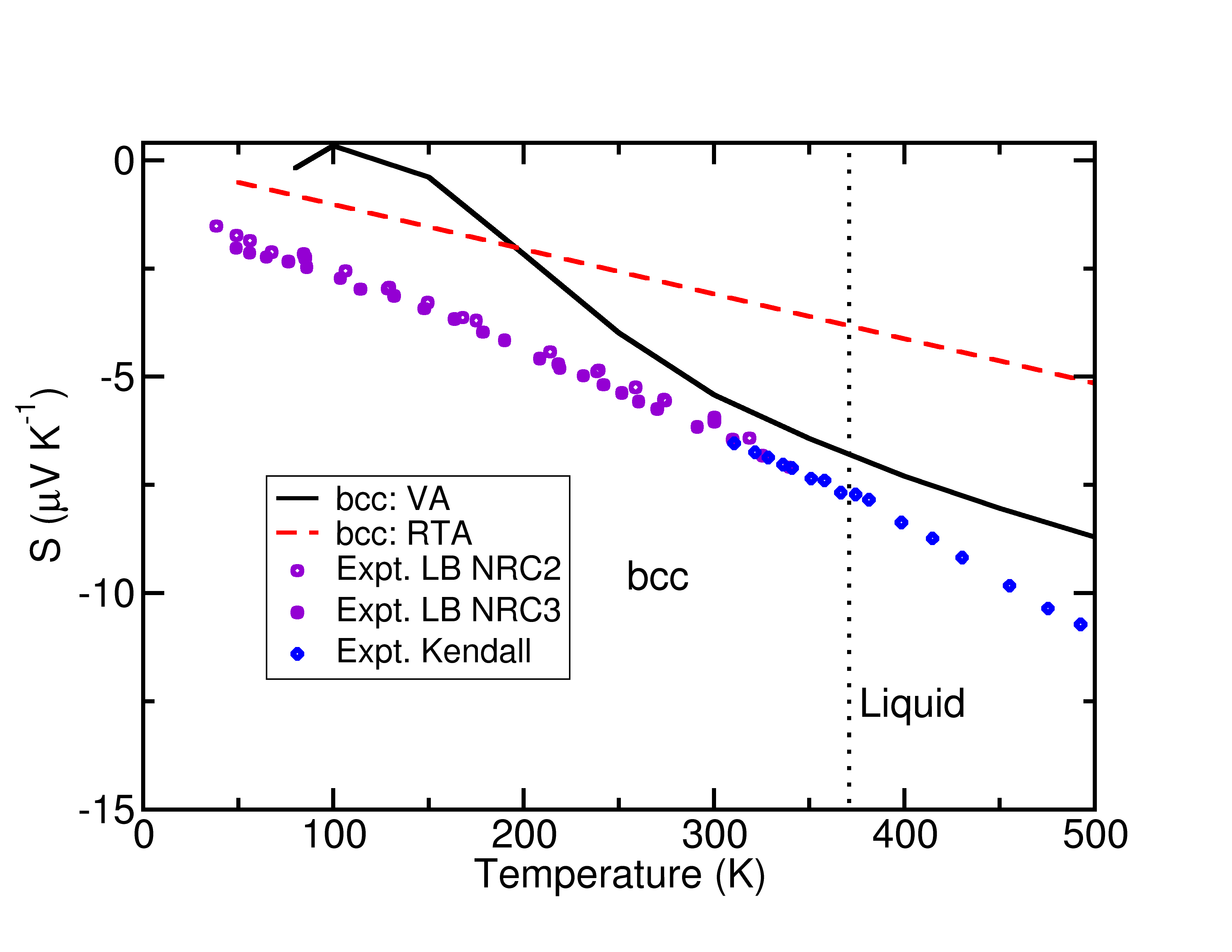}
\caption{(Color online) Seebeck coefficient of Na, as a function of temperature. Black solid line is the calculated $S$ of bcc Na using the variational approach, and the red dashed line denotes the calculated result of bcc Na with constant $\tau$. Discrete points are experimental data from P. Kendall\cite{kendall} and LB(Landolt-B\"{o}rnstein Database)\cite{LBdatabase}. 
The vertical dotted line denotes the temperature of melting point.} \label{fig:sbk_na}
\end{figure}

The calculated $S$, together with the experimental data, are shown in Fig. \ref{fig:sbk_li}. Within the temperature range where the bcc phase is stable, our prediction using the variational approach (VA) agrees very well with the measured $S$, except the older data from Bidwell, which deviates significantly from other experimental values. The excellent agreement on the magnitude and temperature dependence of $S$ also implies that electron-phonon coupling is the main if not sole contribution to the electronic transport properties in bcc Li. This can also be seen from the agreement with measured data\cite{chi1979} in electrical resistivity (See Supplementary Information Fig. 2). On the other hand, $S$ calculated with a constant relaxation time (red dashed line in Fig. \ref{fig:sbk_li}) is negative for all temperatures. This is a clear and qualitative failure of the constant RTA.
In the case of Na (Fig. \ref{fig:sbk_na}), both theoretical predictions are consistent with the experimental sign of $S$, i.e., negative. Comparing to the magnitude of room-temperature $S$ ($\sim$-6 $\mu$V/K) in experiments,\cite{kendall,LBdatabase} the VA result (-5.42 $\mu$V/K) is in good agreement, whereas the constant RTA (-3.09 $\mu$V/K) underestimates by nearly 50\%.

To understand the positive sign of $S$ in Li, we perform a comparative analysis with Na (See Supplimentary Information for details). According to Eq. (\ref{eq:s}), the sign of $S$ is determined by the sign of $Q_{01}$. The sign of $Q_{01}$ comes from the integral of $\alpha_{\text{tr}}^2F$ and $J_{01}$ over the electron energy $\int d\epsilon$. The energy dependence of $J_{01}$ alone favors the negative sign of $S$, for both Li and Na. The different signs of $S$ in Li and Na originate from the energy dependence of $\alpha_{\text{tr}}^2F$. To examine Robinson's hypothesis\cite{robinson1967,robinson1968} that positive $S$ in Li is caused by the unusual energy dependence of electron-phonon interactions, we set the electron-phonon coupling matrix to a constant ($g_{\textbf{kk}'}=1$). A positive value of $S$ is again obtained for Li, and negative for Na. Since the normalized function $F(\textbf{k})$ has a very weak dependence on the energy, the strong energy dependence of $\alpha_{\text{tr}}^2F(\epsilon)$ in Li is due to the integration weights $\delta(\epsilon_{\textbf{k}}-\epsilon)$ (see Supplementary Information Eq. (1)) which is essentially the density of states.

\begin{figure}
\includegraphics[totalheight=0.6\textheight]{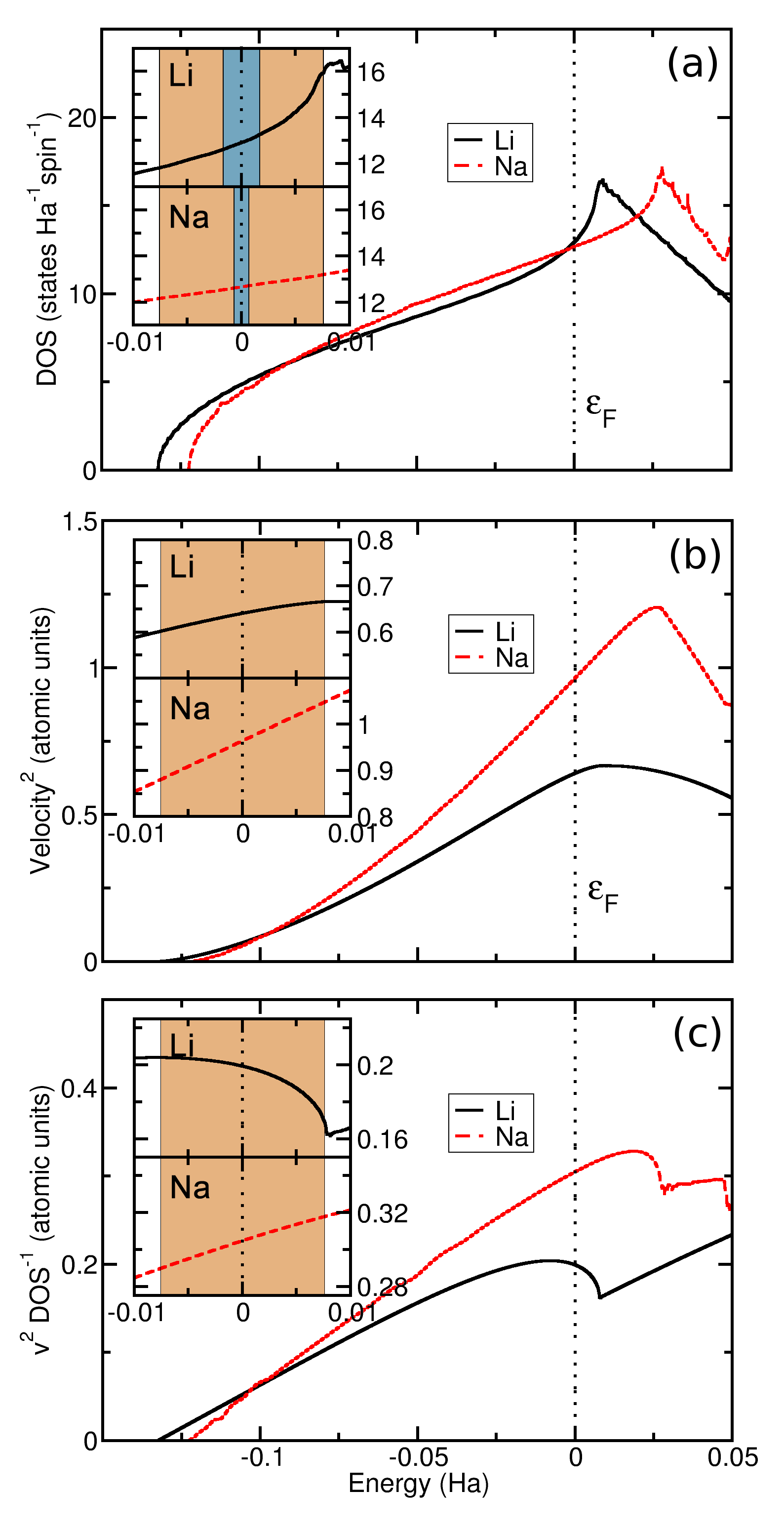}
\caption{(Color online) (a) Density of states, (b) square of the velocity, (c) square of the velocity divided by the corresponding density of states of Li (solid black line) and Na (dashed red line). The vertical dotted line denotes the Fermi energy. The insets show a zoom around $\epsilon_{\text{F}}$. The orange shaded region covers $\epsilon_{\text{F}}\pm 8k_{\text{B}}T$ with $T=300$ K. The blue shaded regions show $\epsilon_{\text{F}}\pm \omega_m$ where $\omega_m$ is the maximum energy for phonons.
} \label{fig:dos}
\end{figure}

We infer that the sign of $S$ is determined by the energy dependence of the electron lifetime. This can also be seen from the RTA point of view. According to Eq. (\ref{eq:sbk}), as $-\partial f/\partial \epsilon$ is positive and symmetric about the Fermi level $\epsilon_{\text{F}}$, the sign of $S$ is determined by the energy dependence of $\sigma(\epsilon)$. Increasing $\sigma(\epsilon)$ yields negative $S$, and vice versa. For a cubic system, without anisotropy, $\sigma(\epsilon)$ consists of $v^2(\epsilon)$ and $\tau(\epsilon)$. We have shown that constant RTA yields negative $S$ for Li (Fig. \ref{fig:sbk_li}), so the energy dependence of $\tau(\epsilon)$ is crucial. 

If the EPC is featureless, the energy dependence of the electron-phonon scattering rate $1/\tau(\epsilon)$ is proportional to the density of states.\cite{ziman1960,lundstrom2009} The electron DOS $N(\epsilon)$ of Li and Na are illustrated in Fig. \ref{fig:dos}(a). As the sign of $S$ does not change with temperature for Li or Na, we choose $T=300$ K for demonstration. In general, a span of $\epsilon_{\text{F}} \pm 8k_{\text{B}}T$ is sufficient to capture the substantial contributions to the transport properties (of $S$, $\rho$, etc). Near the Fermi level, the DOS of Na varies  slowly and does not deviate much from the free-electron description. However, Li exhibits a significantly increasing DOS across $\epsilon_{\text{F}}$, until the band reaches the boundary of the Brillouin zone at the $N$ point. For $v^2(\epsilon)$, i.e., $\sum_{\textbf{k}} v_{x}(\textbf{k})v_{x}(\textbf{k}) \delta(\epsilon-\epsilon_{\textbf{k}})$, Na still shows the free-electron-like linear energy dependence near $\epsilon_{\text{F}}$, but with a much larger slope than Li (Fig. \ref{fig:dos}(b)). The variation of $v^2(\epsilon)$ in Li approaches a plateau just after $\epsilon_{\text{F}}$. Again this behavior deviates qualitatively from the free-electron model, where a linear dependence is expected. Combining these two factors, as shown in Fig. \ref{fig:dos}(c), an increasing $v^2(\epsilon)/N(\epsilon)$ is obtained for Na for the considered energy range around $\epsilon_{\text{F}}$, whereas it is decreasing for Li. At elevated temperatures, although $v^2(\epsilon)/N(\epsilon)$ no longer decreases monotonically for Li due to the larger energy span, the sign of $S$ is unaffected as the major contribution is still from electrons with energies close to the Fermi level.

The relationship between the sign of $S$ and the energy dependence of conductivity can also be seen from the Mott relation, i.e., 

\begin{equation} \label{eq:mott}
  S=-\frac{\pi^2 k^2_{\text{B}}T}{3e} \left[ \frac{1}{\sigma} \frac{d \sigma(\epsilon)}{d\epsilon} \right]_{\epsilon=\epsilon_{\text{F}}}
\end{equation} 

As a qualitative estimation, if Drude's formula $\sigma=ne^2\tau/m^*$ is adopted for the conductivity and again the relaxation time $\tau$ is taken to be inversely proportional to the DOS, the energy dependencies from the charge carrier density and $\tau$ are approximately balanced out, so that $\sigma$ has the same energy dependence as $1/m^*$. For Li, as implied by the DOS in Fig. \ref{fig:dos}(a), the band becomes flattened around the Fermi energy which corresponds to an increasing effective mass. Consequently, $\sigma$ decreases with energy and yields the positive sign of $S$.

\begin{figure}
\includegraphics[width=1.0\columnwidth]{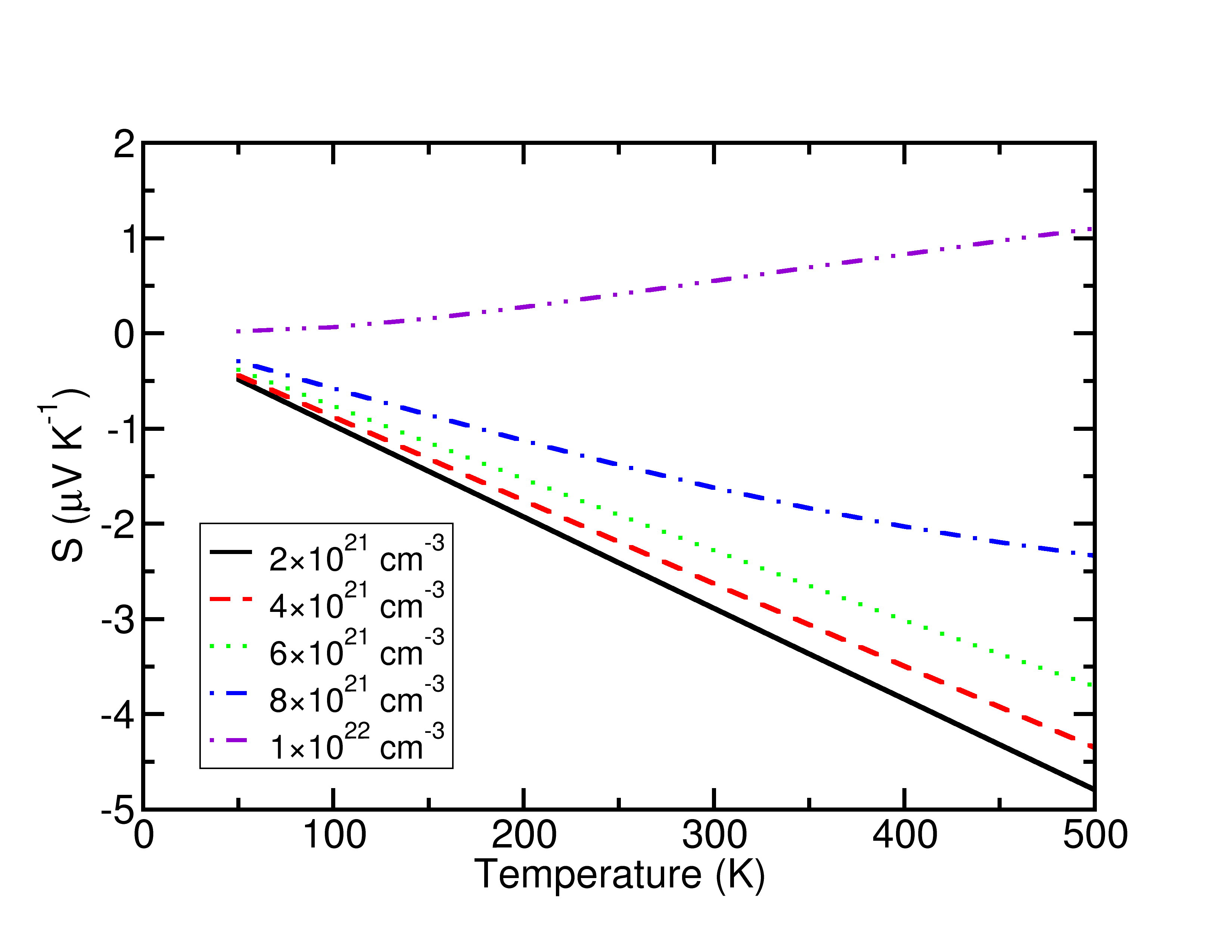}
\caption{(Color online) Calculated Seebeck coefficient of electron-doped Na using RTA and $1/\tau(\epsilon) \propto N(\epsilon)$, as a function of temperature at several doping concentrations.} \label{fig:n_na}
\end{figure}

We now turn to the possibility of doping-induced sign changes in $S$. If Na is electron doped, using the relaxation time approach and the qualitative relation between $\tau$ and DOS, we predict that the sign of $S$ changes from negative to positive with a concentration $\sim$ 1$\times$10$^{22}$ cm$^{-3}$ (0.358 e$^-$/unit cell), cf. Fig. \ref{fig:n_na}. This change of sign is confirmed in the VA, for slightly higher doping levels but with a much stronger amplitude: at 300 K, $S=0.55$ $\mu$V/K from RTA while $S=5.53$ $\mu$V/K using VA.

Clearly the proportionality between the scattering rate and DOS is qualitative, and works for simple systems. 
The $\tau(\epsilon)$ model fails in particular for Fermi surfaces not entirely within the first Brillouin zone. As an example, if extra electrons are added to Li, e.g. in Mg$_x$Li$_{1-x}$ alloy,\cite{nayeb1984} the RTA with the model $\tau(\epsilon)$ yields a change of sign of $S$ from positive to negative at an extra electron concentration of about 8$\times$10$^{21}$ cm$^{-3}$ (Mg$_{0.154}$Li$_{0.846}$). However, the VA calculated $S$ does not change sign, at least up to an added carrier concentration of 4$\times$10$^{22}$ cm$^{-3}$ (Mg$_{0.771}$Li$_{0.229}$ which is beyond the wide range of BCC structure for the binary alloy). 
When the Fermi surface reaches the BZ boundary, the distortions allow additional electron phonon scattering that will change the scattering rate. Similar failures of the RTA with either constant or DOS-related $\tau$ are found in Cu, Ag, and Au, where the model $\tau(\epsilon)$ still gives negative $S$. The positive $S$ in these group-IB metals is more complex than in Li, combining a distorted FS with non-trivial electron-phonon interactions, as was proposed by Robinson \cite{robinson1967,robinson1968}. Fully first-principles calculations are underway to elucidate the precise mechanism.

In summary, we have calculated the first fully \emph{ab initio} Seebeck coefficient, using a variational solution to the Boltzmann transport equation. Our calculated Seebeck coefficients of Li and Na are in good agreement with experimental data, whereas the commonly used constant relaxation time approximation fails qualitatively for Li. By comparison between Li and Na, a detailed analysis reveals that the sign of $S$ is determined by the energy dependence of the electron lifetime (generically proportional to the inverse of the electronic DOS), whereas the quantitative influence of the electron-phonon interaction is not important. In Li, the DOS around the Fermi energy deviates considerably from the free-electron model, and our analysis contradicts Robinson's earlier explanations based on exotic energy variations of the electron-phonon coupling. The possibility of tailoring the sign of the Seebeck coefficient through electronic band engineering opens many pathways to improved thermoelectric devices. In more complex cases electron-phonon coupling effects will probably be as important as the purely electronic effects for the net variation of the electron lifetime; both are included seamlessly and on the same footing in the present formalism.

\begin{acknowledgments}
The authors thank Y Pouillon for essential build system work on the ABINIT package. We acknowledge an A.R.C. grant (TheMoTherm 10/15-03) and a ULg starting grant (cr\'edit d'impulsion ``Etude ab initio des transitions de phase ˆ haute temp\'erature''), both from the Communaut\'e Fran\c{c}aise de Belgique. Computer time was made available by PRACE-2IP on Huygens and Hector (EU FP7 Grant No. RI-283493), CECI, CISM-UCLouvain, and SEGI-ULg.
\end{acknowledgments}

\bibliography{li.bib}

\end{document}